\documentclass[prl,twocolumn,english,superscriptaddress,floatfix,longbibliography,aps]{revtex4-1}
\usepackage{float}
\usepackage{graphicx}
\usepackage{amsmath, amssymb, amsfonts}
\usepackage{amsmath, amssymb}
\usepackage{verbatim}
\usepackage{lipsum}
\usepackage{xcolor}
\usepackage{hyperref}
\hypersetup{colorlinks = true,urlcolor = blue}

\def\dbar{{\delta}}
\def\tr{\mathrm{tr}}

\def\be{\begin{equation}}
\def\ee{\end{equation}}
\def\bea{\begin{eqnarray}}
\def\eea{\end{eqnarray}}

\newcommand{\comm}[1]{{\color{black} #1}}
\newcommand{\addkm}[1]{{\color{black} #1}}

\newcommand{\pmh}[1]{{\color{black} #1}}

\newcommand{\aleadd}[1]{{\color{black} #1}}
\newcommand{\el}[1]{{\color{black} #1}}
\newcommand{\kwm}[1]{{\color{black} #1}}
\newcommand{\ar}[1]{{\color{black} #1}}
 
\begin{document}

\title{Heat and work along individual trajectories of a quantum bit}

\author{M. Naghiloo}
\affiliation{Department of Physics, Washington University, St.\ Louis, Missouri 63130}
\author{D. Tan}
\affiliation{Department of Physics, Washington University, St.\ Louis, Missouri 63130}
\affiliation{Shenzhen Institute for Quantum Science and Engineering and Department of Physics, Southern University of Science and Technology, Shenzhen 518055, People's Republic of China}
\author{P. M. Harrington}
\affiliation{Department of Physics, Washington University, St.\ Louis, Missouri 63130}
\author{J. J. Alonso}
\affiliation{Department of Physics, Friedrich-Alexander-Universit\"at Erlangen-N\"urnberg, D-91058 Erlangen, Germany}
\author{E. Lutz}
 \affiliation{Institute for Theoretical Physics I, University of Stuttgart, D-70550 Stuttgart, Germany}
\author{A. Romito}
\affiliation{Department of Physics, Lancaster University, Lancaster LA1 4YB, United Kingdom}
\author{K. W. Murch}
\affiliation{Department of Physics, Washington University, St.\ Louis, Missouri 63130}
\affiliation{Institute for Materials Science and Engineering, St.\ Louis, Missouri 63130}

\date{\today}

\begin{abstract}
We use a near quantum limited detector to experimentally track individual quantum state trajectories of a driven qubit formed by the hybridization of a waveguide cavity and a transmon circuit. For each measured quantum coherent trajectory, we separately identify energy changes of the qubit as heat and work, and verify the first law of thermodynamics for an open quantum system. We further establish the consistency of these results by comparison with the master equation approach and the two-projective-measurement scheme, both for open and closed dynamics, with the help of a  quantum feedback loop that compensates for the exchanged heat and effectively isolates the qubit.
\end{abstract}

\maketitle

Continuous measurement of a quantum bit can be used to track individual trajectories of its state. Due to the intrinsic quantum fluctuations of a detector, measurement is an inherently stochastic process \cite{jac14}. If a quantum system starts in a given state, then by accurately monitoring the fluctuations of the detector, it is possible to reconstruct single quantum trajectories, which describe the evolution of the quantum state conditioned to the measurement outcome \cite{jac14}. The idea of quantum trajectories made its transition from a theoretical tool (unraveling)  to simulate open quantum systems \cite{car93} to a physically accessible quantity with the experimental ability of tracking these trajectories in optical \cite{gue07,say11} and more recently in solid state \cite{murc13,webe14} systems. Continuous monitoring of superconducting qubits has, for example, enabled continuous feedback control \cite{vija12,blo14,lan14}, the determination of weak values \cite{gro13,cam14,tan15}, and the production of deterministic entanglement \cite{ris13,roc14}. In view of their ability to combine quantum trajectory monitoring with external unitary driving, these superconducting devices additionally offer a unique platform to explore energy exchanges and thermodynamics along single quantum trajectories.

The laws of thermodynamics classify energy changes for macroscopic systems as work performed by external driving and heat exchanged with the environment \cite{pip66}. In past decades,  these principles have been successfully extended to the level of classical trajectories  to account for thermal fluctuations \cite{jar11}. By providing  a theoretical and experimental framework for determining work and heat along individual trajectories, stochastic thermodynamics  has  paved the way for the study of the energetics of microscopic systems, from colloidal particles to enzymes and molecular motors \cite{sei12,cil13}.
The further generalization of thermodynamics to include quantum fluctuations faces unique challenges, ranging from the proper identification of  heat and work to the clarification of the role of coherence \cite{gal16,deff16,kamm16,cott17}. 
\pmh{Quantum heat is commonly associated with the nonunitary part of the dynamics \cite{ali79,spo79,kos84}, carrying over the classical notion of  energy exchanged with the surroundings. This definition has  recently been extended to the level of single discrete quantum jumps \cite{bre03,leg13,hekk13,gon16,horo12,hor13} and to individual continuous quantum trajectories \cite{alon16,elou17}. Other definitions of quantum work and heat have been put forward, for instance based on the single shot approach \cite{abe13,hal15} or quantum resource theory \cite{bra13,bra15}. This diversity of theoretical approaches emphasizes the crucial importance of an experimental study.}

We here report the measurement of work and heat associated with unitary and non-unitatry dynamics along single quantum trajectories of a superconducting qubit. The qubit evolves under continuous unitary evolution and is only weakly coupled to the detector. As a result, information about its state may be inferred from the measured signal without projecting it into eigenstates. This system might thus generically be in coherent superpositions of energy eigenstates. We show that the measured heat and work are consistent with the first law and prove the agreement with both the two-projective-measurement (TPM) scheme \cite{talk07} and the master equation approach  \cite{ali79,spo79,kos84}. We finally establish the correspondence with the TPM work in the unitary limit by employing a phase-locking quantum feedback loop that effectively compensates for the heat.

%\ar{Here we report the measurement of work and heat associated with unitary and non-unitatry dynamics along single quantum trajectories of a superconducting qubit. The qubit evolves under continuous unitary evolution and weak measurement, so that it remains in coherent superposition of energy eigenstates.}
%% Here we \el{demonstrate} an experimental protocol to measure heat and work along single quantum trajectories of a superconducting qubit evolving under continuous unitary evolution and measurement. \er{The qubit is only weakly coupled to the detector and information about its  state may be inferred from the measured signal without projecting it into eigenstates. This system might thus generically be in coherent superpositions of energy eigenstates.}
% We show that the measured heat and work are consistent with the first law and \el{prove}  the agreement with both the two-projective-measurement (TPM) scheme \cite{talk07} and with \ar{the average heat and work from} the master equation approach \cite{ali79,spo79,kos84}. We finally establish the \el{correspondence} with the TPM work in the unitary limit by employing a phase-locking quantum feedback loop that effectively \addkm{compensates} for the \pmh{heat}.

\textit{Heat and work along quantum trajectories.}
In macroscopic thermodynamics, work performed on a thermally isolated system is defined as the variation of internal energy, $W=\Delta U$ \cite{pip66}.  According to the first law, heat is given by the difference, $Q =\Delta U -  W$, for systems that are not thermally isolated \cite{pip66}. Thermal isolation is thus essential to distinguish heat  from work.  {At the quantum level, identifying heat and work  is more involved, because quantum systems do not necessarily occupy definite} {energy} states.  Energy changes are usually defined  in terms of transition probabilities between energy eigenstates in the so-called two-projective-measurement (TPM) scheme \cite{talk07}. For a driven quantum system described by the Hamiltonian $H_t$, the  distribution of  the total energy variation $\Delta U$ is thus given by \cite{talk07},
\begin{equation}\label{eq:pu}
P(\Delta U)  =  \sum_{m,n} P_{m,n}^{\tau} P_{n}^{0} \delta[\Delta U - (E^\tau_m-E^0_n)],
\end{equation}
where $P_{n}^{0}$ denote the initial occupation probabilities, $P_{m,n}^{\tau}$ are the transition probabilities  between   initial and final eigenvalues $E^0_n$ and $E^\tau_m$ of  $H_t$,  and $\tau$ is the duration of the {driving} protocol. 
This relation has been used to {experimentally} determine the work distribution in \kwm{closed} quantum systems such as NMR, trapped ion, and cold atom systems  \cite{bata2014,shuo2015,cer17}, for which $\Delta U= W$.  

However, in open quantum systems, the total energy change $\Delta U$ cannot, in general, be uniquely separated into heat and work \cite{cam11} and several definitions  have been proposed \aleadd{\cite{alon16,elou17,ali79,spo79,kos84,bre03,leg13,hekk13,gon16,horo12,hor13,abe13,hal15,bra13,bra15}}.  \ar{Open quantum systems can be} described with density operator $\rho_t$ with evolution given by a quantum master equation \cite{bre02},
\begin{equation}
\label{11}
\frac{d\rho_t}{dt} = -\frac{i}{\hbar}[H_t, \rho_t]+ {\cal L} \rho_t,
\end{equation}
where ${\cal L}$ is a Lindblad dissipator. In this case, the first law has  \el{been} written in the usual form, $\Delta \bar U = \bar Q+ \bar W$, with \cite{ali79,spo79,kos84},
\begin{equation}
\label{12}
\bar Q = \int_0^\tau dt\,\text{tr}\left[\frac{d\rho_t}{dt}H_t\right ], \,\,\, \bar W= \int_0^\tau dt\, \text{tr}\left[\rho_t\frac{dH_t}{dt}\right].
\end{equation}
As in classical thermodynamics, $\bar Q$ is  the energy supplied  to the system by the environment and $\bar W$ the work done \kwm{by external} driving. The above definition of quantum work has been originally introduced by Pusz and Woronowicz in a $C^*$-algebraic context \cite{pus78} and recently  applied to  individual discrete quantum jumps \cite{bre03,leg13,hekk13,gon16}.

In our experiment, we  examine how quantum heat and work can be consistently identified  for systems whose environment consists of a continuously coupled quantum limited detector, an effectively zero temperature \ar{reservoir} \cite{jac14}. \kwm{The ability to track quantum state trajectories enables energy changes to be decomposed separately into heat and work components} \cite{alon16,elou17}. {The starting point of our analysis is that} {the quantum state} evolution consists of both a unitary part and, because of the continuous monitoring, an additional nonunitary  component: the former is again identified as work, the latter as heat, {in analogy to macroscopic thermodynamics} \cite{alon16,elou17}.
Specifically, for an infinitesimal time interval $dt$, {a change of the conditional density operator for a single trajectory may be written as} $d \tilde{\rho}_t= \delta \mathbb{W}[\tilde \rho_t] dt+ \delta \mathbb{Q}[\tilde \rho_t]dt$, where  $\delta \mathbb{W}[\tilde \rho_t] $ and $\delta \mathbb{Q}[\tilde \rho_t] $ are  \comm{superoperators} associated with the respective unitary and nonunitary dynamics \cite{alon16}. The {tilde} {here marks \el{quantities that are evaluated in different realizations of the experiment, as opposed to quantities averaged over the possible trajectories.}}
 The first law  along a single quantum trajectory $\tilde \rho_t$ then reads $d \tilde U=\delta \tilde W + \delta \tilde Q$, with $\delta\tilde W =\tr[\tilde\rho_{t-dt}dH_t]$ and $\delta\tilde Q=\tr[H_{t}d\tilde{\rho}_t]$ \cite{sm}. 
 When integrated over time, the first law takes the form,
\kwm{\begin{equation}
\label{2}
\Delta \tilde U=\int_0^\tau  \frac{d \tilde U}{dt}\, dt = \int_0^\tau  \frac{\delta \tilde W}{dt} \, dt + \int_0^\tau  \frac{\delta \tilde Q}{dt}  \, dt,
\end{equation} }
for each  quantum trajectory. Equation \eqref{2} is a quantum extension of the first law of stochastic thermodynamics. {It relates the   average change of energy $\Delta  \tilde U $ with the path-dependent heat $\tilde Q$ and work $\tilde W$.} Similarly, we may distinguish quantum heat and work contributions to changes of the transition probabilities \cite{alon16},
\be
\label{21}
 d\tilde P_{m,n} =\delta\tilde P^W_{m,n}+\delta\tilde  P^Q_{m,n},
 \ee
 along single  quantum trajectories \cite{sm}.  

\kwm{The consistency of the decompositions \eqref{2} and \eqref{21} may be established in three independent  ways: \textit{(i)} the total energy change along a trajectory, $\Delta \tilde U = \sum d \tilde U$, and the total transition probability, $\tilde P_{nm} = \sum d\tilde P_{nm}$, may be compared to the TPM approach \cite{talk07}, \textit{(ii)} the stochastic heat and work contributions \eqref{2} may be compared to the mean quantities \eqref{12} after averaging over stochastic and quantum fluctuations, and \textit{(iii)} finally, the work \eqref{2} along a trajectory  may be directly compared to  the TPM result \eqref{eq:pu} in the unitary limit when  heat vanishes. In that case,  $\Delta \tilde U = \Delta U = E_m^\tau-E_n^0=W$ \cite{sm}. }%Note that Eq.~\eqref{2} allows   for initial superposition of states, while Eq.~\eqref{eq:pu} does not. 

\begin{figure*}[!htb]
\begin{center}
\includegraphics[width =.8\textwidth]{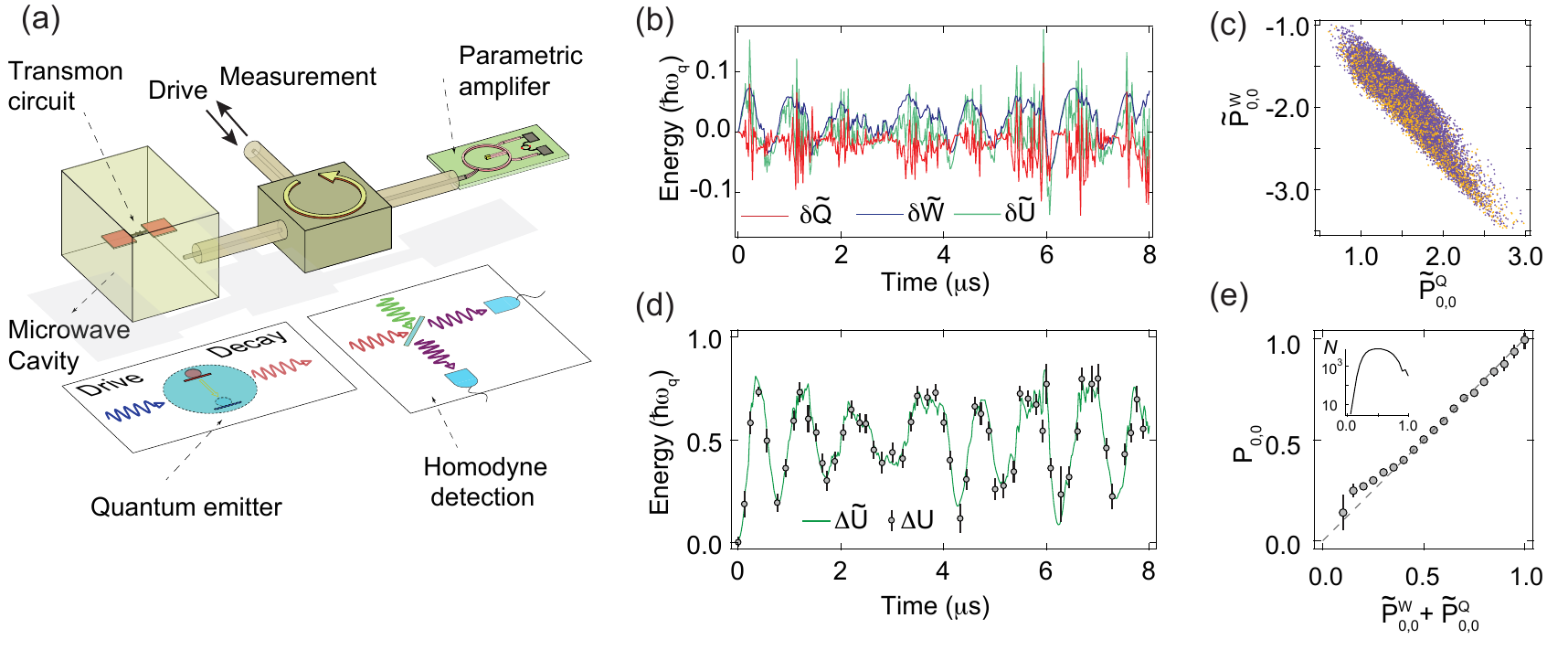}
\caption{\small  {Evaluating heat and work along single quantum trajectories.}  
(a), Schematic of the qubit system, drive, and homodyne detection. 
(b), Work (blue), heat (red) and energy (green) along a single trajectory. \aleadd{The discrete timestep resolution is $\delta t= 20$ ns, the smallest compatible with the detection bandwidth}
(c), A scatter plot of  final work and heat contributions to the  $P_{0,0}$ transition probability for an ensemble of $\sim 10^4$ experimental protocols of duration $2\ \mu$s.  Each experimental sequence terminates with a projective measurement and the color of the points indicate the outcome of this measurement (orange: $m=1$, purple: $m=0$). \addkm{The heat and work contributions are not necessarily bounded, but their sum is limited to [-1,0] as expected.}
(d), The total energy along a single quantum trajectory (green) compared to the total energy as determined from an ensemble of projective measurements at each time point (circles). The error bars indicate the standard error of the mean.
(e), Projective measurements binned and averaged according to the sum of the work and heat contributions $\tilde{P}_{0,0}^W+\tilde{P}_{0,0}^Q$.  The error bars indicate the standard error of the mean based on the number of occurrences ($N$) for each value of $\tilde{P}_{0,0}^W+\tilde{P}_{0,0}^Q$ (inset). }
\label{fig:fig1}
\end{center}
\end{figure*}

%%%%%%%%%%%%%%%%%%%%%%%%%%%%%%%%%%%%%%%%%
%EXPERIMENTAL SYSTEM
%%%%%%%%%%%%%%%%%%%%%%%%%%%%%%%%%%%%%%%%%

\textit{Experimental set-up}.
The qubit is realized by the near-resonant interaction of a transmon circuit   \cite{koch07}
 and a three dimensional aluminum cavity  \cite{paik113D} \kwm{capacitivley} coupled to a $50\  \Omega$ transmission line. Resonant coupling between the circuit and cavity results in an effective qubit  which is described by the Hamiltonian,  $H_\mathrm{q}= -\hbar \omega_\mathrm{q}\sigma_z/{2}$,  and depicted in Figure~\ref{fig:fig1}a.  The radiative interaction between the qubit and  transmission line is given by the interaction Hamiltonian, $H_\mathrm{int}=\hbar\gamma(a \sigma_+ + a^{\dagger} \sigma_ -)$, where $\gamma$ is the coupling rate between the electromagnetic field mode corresponding to $a$ ($a^\dagger$), the annihilation (creation) operator, and the qubit state transitions denoted by $\sigma_+$ ($\sigma_-$), the raising (lowering) ladder operator for the qubit. \kwm{By virtue of this interaction Hamiltonian, a homodyne measurement along an arbitrary quadrature of the quantized electromagnetic field of the transmission line, $a e^{-i\varphi} + a^{\dagger} e^{+i\varphi}$, results in weak measurement along the corresponding dipole of the qubit, $\sigma_+ e^{-i\varphi} + \sigma_- e^{+i\varphi}$   \cite{nagh16}. In order to perform work on the qubit, we introduce a  classical time-dependent field described by the Hamiltonian $H_{\mathrm{R}} =  \hbar \Omega_\mathrm{R}  \sigma_y \cos(\omega_\mathrm{q} t + \varphi)$, %\comm{[ $H_{\mathrm{R}} = - \Omega_\mathrm{R}/2~  \sigma_y$ in the rotating frame)]}
where $\omega_\mathrm{q}$ is the resonance frequency of the qubit and $\Omega_\mathrm{R}$ is the Rabi drive frequency.}
 
Homodyne monitoring is performed with a  Josephson parametric amplifier \cite{cast08,hatr11para}   operated in phase-sensitive mode. We adjust the homodyne detection quadrature such that the homodyne signal $dV_t$ obtained over the time interval $(t, t+dt)$ provides an indirect signature  \cite{jord15}     of the real part of $\sigma_- = (\sigma_x+i\sigma_y)/2$. The detector signal  is given by $d V_t = \sqrt{\eta}\gamma \langle \sigma_x \rangle dt + \sqrt{\gamma} dX_t$, where $\eta$ is the quantum efficiency of the homodyne detection, $\gamma$ is the radiative decay rate, and $dX_t$ is a zero-mean Gaussian random variable with variance $dt$.

The qubit evolution, given both driven evolution $H_\mathrm{R}$ and homodyne measurement results $dV_t$, is described in the rotating frame by the stochastic master equation  \cite{bolu14},
\begin{align}\label{eq:sme}
\begin{split}
d \tilde \rho_t &=  - \frac{i}{\hbar} [H_R, \tilde \rho_t]\,dt + \gamma \mathcal{D}[\sigma_-] \tilde \rho_t\,dt\\
&\hspace{3.5cm}+ \sqrt{\eta \gamma } \mathcal{H}[\sigma_- dX_t] \tilde \rho_t ,
\end{split} 
\end{align}
where $ \mathcal{D}[\sigma_-] \tilde \rho = \sigma_- \tilde \rho \sigma_+ -\frac{1}{2} (\sigma_+ \sigma_- \tilde \rho + \tilde \rho  \sigma_+ \sigma_- )$ and $ \mathcal{H}[O]  \tilde \rho = O \tilde\rho + \tilde \rho O^\dagger - \mathrm{tr}[(O+O^\dagger)\tilde \rho]\tilde \rho$ are the dissipation and jump superoperators, respectively.  By taking the ensemble average, Eq.~\eqref{eq:sme} reduces to a master equation  of the  form \eqref{11} with  dissipator ${\cal L}\rho_t = \gamma {\cal D}[\sigma_-]\rho_t$, which describes the coupling to a zero-temperature reservoir \cite{jac14}.

%%%%%%%%%%%%%%%%%%%%%%%%%%%%%%%%%%%%%%%%%
%HEAT AND WORK
%%%%%%%%%%%%%%%%%%%%%%%%%%%%%%%%%%%%%%%%%

We next introduce the experimental protocols to determine the stochastic heat and work contributions.
 We  identify the  instantaneous work contribution $\dbar\mathbb{W}[\tilde \rho_t]$ with the  first (unitary) term in Eq.~(\ref{eq:sme}), while the  instantaneous heat contribution $\dbar \mathbb{Q}[\tilde \rho_t]$ is associated with the latter two (nonunitary) terms.  \addkm{Although the system could, in general, exchange energy with the detector in the form of heat or work, the homodyne measurement in our experiment only induces a zero-mean stochastic back-action, which guarantees no extra work is done by the detection process.}

Having access to the instantaneous heat and work contributions \addkm{from} an individual  quantum trajectory, we now verify the  first law    in the form of Eqs.~(\ref{2}) and \eqref{21}. For this, we initialize the qubit in the eigenstate $n$,  and then drive the qubit while collecting the homodyne measurement signal. Figure~\ref{fig:fig1}b shows the path-dependent instantaneous heat and work  contributions,  $\dbar \tilde {Q}$ and $\dbar \tilde {W}$, and the corresponding changes in internal energy $d\tilde U$  for a single  trajectory originating in $n=0$.  After time $\tau$, we projectively measure   \cite{reed10} the qubit in state $m$ and then repeat the experiment several times. In Figure~\ref{fig:fig1}c  we show a scatter plot of the calculated heat and work contributions {to the transition probabilities}, $\tilde{P}_{m,n}^Q$ and $\tilde{P}_{m,n}^W$, for $\tau = 2\ \mu$s. Each single quantum trajectory exhibits different heat and work contributions,  highlighting the stochastic nature of its quantum evolution. \kwm{Using individual heat and work trajectories we now address the consistency of these decompositions in three independent ways.}

\kwm{\textit{(i) Total energy change}---}In order to establish the consistency of these results with the  TPM scheme \cite{talk07}, we first show in Figure~\ref{fig:fig1}d the path-dependent total energy variation $\Delta \tilde U =\sum \delta \tilde U$ for a single trajectory and the path-independent total energy change $\Delta U =(\hbar \omega_q) P_{1,0}^\tau$ obtained via projective measurements performed at various intermediate times \cite{sm}. We find that the path-independent energy changes are in excellent agreement with the energy changes along a single quantum trajectory.  In Figure~\ref{fig:fig1}e we further compare the  path-independent transition probability  $P_{0,0}$ to the sum of the path-dependent work and heat contributions, $\tilde P_{0,0}^W + \tilde P_{0,0}^Q$, for experiments of variable duration $\tau = [0,8]\ \mu$s. We again observe very good agreement.

\begin{figure}[t]
\begin{center}
\includegraphics[width =.5\textwidth]{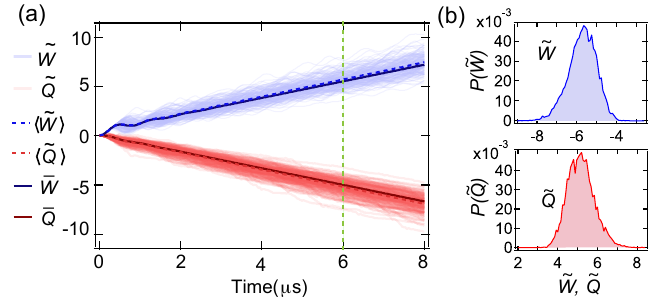}
\end{center}
\caption{\kwm{Comparison \el{of} stochastic and average heat and work quantities. (a), Individual heat and work trajectories $\tilde{Q}$, $\tilde{W}$ are displayed as transparent red and blue traces. The mean of these individual trajectories $\langle \tilde{Q}\rangle$, and $\langle \tilde{W} \rangle$ are displayed as dashed lines which are in good agreement with the mean values \el{from the master equation}, $\bar Q$ and  $\bar W$, \el{Eq.~\eqref{12}}, solid lines. (b),  Distributions of $\tilde{Q}$ \el{and} $\tilde{W}$ at evolution time $\tau=6\ \mu$s.}}   \label{fig:distributions}
\end{figure}

\kwm{\textit{(ii) Correspondence with master equation definitions}---}Figure \ref{fig:distributions} displays the time evolution of the heat $\tilde Q$ and work $\tilde W$ along single trajectories, as well as \kwm{their} respective mean values.  The ensemble average of the individual work $\langle \tilde{W}\rangle$ and heat $\langle \tilde{Q} \rangle$ trajectories agrees well with the the averaged values, $\bar Q$ and  $\bar W$, Eq.~\eqref{12}, thus recovering the expression by Pusz and Woronowicz  \cite{pus78} \kwm{at the level of unraveled quantum trajectories}.  In addition, the individual trajectories allows examination of the heat and work distributions (Fig.~\ref{fig:distributions}b) at each timestep.  

%%%%%%%%%%%%%%%%%%%%%%%%%%%%%%
%% FEEDBACK%%%%
%%%%%%%%%%%%%%%%%%%%%%%%%%%%%%

\kwm{\textit{(iii) The unitary limit}---}We finally \kwm{show} \el{correspondence} of the quantum trajectory work $\tilde W$ and the TPM work, $W = E_m^\tau -E_n^0$, for a single realization by experimentally isolating the system with a quantum feedback loop \cite{jac14}. 
\kwm{The essence of feedback is to compensate for the effect of the detector by adjusting the Hamiltonian at each timestep, $\dbar\mathbb{Q}[\tilde \rho_t]$, thus making the system effectively closed}. The dynamics of the system is then simply described by unitary evolution where only the work $\dbar\mathbb{W}[\tilde \rho_t]$ contributes \kwm{to changes in the state}. In order to implement feedback, we adapt the phase-locked loop protocol introduced in Ref.~\cite{vija12}.  This is achieved by multiplying the homodyne measurement signal with a reference oscillator of the form $A[\sin(\Omega_\mathrm{R} t+\phi)+B]$ yielding a feedback control, $\Omega_\mathrm{F}  =  \sqrt{\eta} (\cos(\Omega t + \phi)-1) dV_t/dt$, that modulates the Rabi frequency of the qubit drive. The detector heat exchange is eliminated by applying additional work, $\dbar\mathbb{W}_F[\tilde \rho_t] =  ({i}/{\hbar}) [\hbar \Omega_\mathrm{F} \sigma_y\cos(\omega_q t + \phi), \rho_t]$.

  \begin{figure}[!htb]
\begin{center}
\includegraphics[width =.45\textwidth]{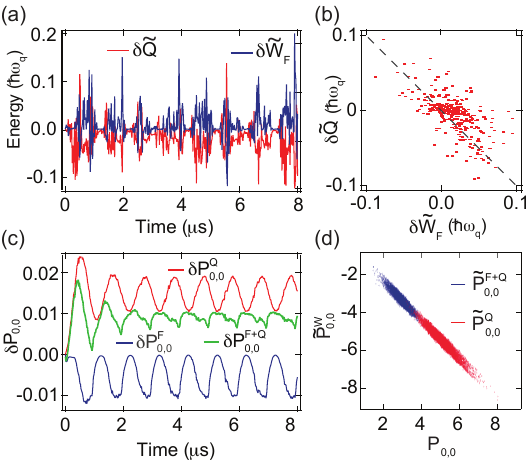}
\caption{\small {Quantum feedback loop.} (a), Instantaneous heat and feedback work along a single trajectory. \comm{The feedback work has been time shifted by $20$ ns to account for the time delay in the feedback circuit.} The anti-correlation ($r=-0.68$) of  heat and feedback work is evident in the scatter plot (b). (c), Average of the instantaneous contribution of heat and feedback to the transition probability for $10^4$ experimental iterations. (d), Parametric plot of $\tilde{P}_{0,0}^{W}$ versus $\tilde{P}_{0,0}^{Q}$ (red) and $\tilde{P}_{0,0}^{F+Q}$ (blue) \el{showing} how the feedback cancels the \el{heat}, narrowing and shifting the distribution toward zero for $\tau = 6 \ \mu$s.}   \label{fig:fig2}
\end{center}
\end{figure}
  
Figure~\ref{fig:fig2}\el{a} shows the instantaneous feedback work, \comm{$\dbar \tilde W_\mathrm{F} =\hbar \omega_\mathrm{q}\mathrm{tr}\left[\Pi_{m=1} \dbar\mathbb{W}_\mathrm{F}[\tilde \rho_t]\right]dt$ (\el{with} $\Pi_m$ the projector onto eigenstate $m$), together with the corresponding instantaneous heat, $\dbar \tilde Q =  \hbar \omega_\mathrm{q}\mathrm{tr}\left[\Pi_{m=1} \dbar\mathbb{Q}[\tilde \rho_t]\right] dt$,} %$\dbar \tilde Q = \mathrm{tr}\left[H_t \dbar\mathbb{Q}[\tilde \rho_t]\right] dt$,
 along a  \el{trajectory} for a quantum efficiency of 35$\%$.  We observe that the feedback partially cancels the heat at each point in time. The anti-correlation between the instantaneous feedback and heat contributions depicted in Figure~\ref{fig:fig2}\el{b} confirms that the feedback loop compensates for exchanged heat at each timestep. In addition, by averaging the instantaneous heat and work contributions to the transition probability over many iterations of the experiment (Fig.~\ref{fig:fig2}c), we  clearly see how feedback works toward canceling the heat  on average. Similarly, at the level of single trajectories, the total transition probability may  be  \el{written} as 
$\tilde{P}_{m,n}^\tau = \tilde{P}_{m,n}^{W} +   \tilde{P}_{m,n}^{Q}+  \tilde{P}_{m,n}^{F}$,  with the work contribution from feedback $\tilde{P}_{m,n}^{F}$. Figure~\ref{fig:fig2}d shows the transition probabilities $\tilde{P}_{0,0}^{W}$ versus  $ \tilde{P}_{0,0}^{Q} + \tilde{P}_{0,0}^{F}$. By comparing the transition probabilities with and without feedback, we observe a significantly reduced heat contribution.

% In this limit, the experimental test would reproduce previous results \cite{bata2014,shuo2015,cer17}  for closed quantum systems.  

\begin{figure}[htb]
\begin{center}
\includegraphics[width =.45\textwidth]{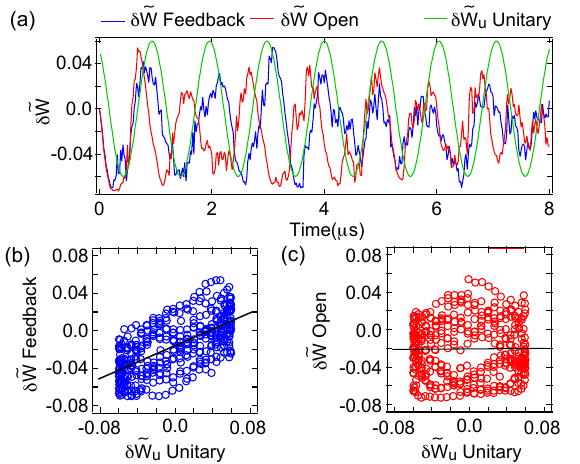}
\end{center}
\caption{Work along trajectories with and without feedback\el{.} (a), the instantaneous work $\delta \tilde{W}$ along a single  trajectory in the presence of feedback (blue) and the open loop configuration (no feedback) (red) is compared to the calculated instantaneous work expected for pure unitary evolution (green). (b, c), The correlation between the instantaneous work $\delta \tilde{W}$ and the work for a unitary evolution along a \el{quantum} trajectory.  The linear regression fit (black lines) show correlation (slope 0.437) when the feedback loop is employed, and no correlation in the open loop configuration (slope 0.006).}    \label{fig:workandunitary}
\end{figure}

In the presence of the quantum feedback loop we can decompose the instantaneous work along trajectories into work imparted by the feedback and work associated with the driving protocol, $\delta \tilde{W}$.  In the absence of the feedback loop, the quantum dynamics of the qubit are given by  work $\delta \mathbb{W}[\tilde \rho_t] $ and heat $\delta \mathbb{Q}[\tilde \rho_t] $ superoperators; the heat changes the state, causing the observed $\delta \tilde{W}$ to differ from the case of closed unitary evolution, $\delta \tilde{W}_\mathrm{u}$.  With the feedback loop, the heat contribution is compensated at each timestep causing the instantaneous work $\delta \tilde{W}$ to match the expected unitary work $\ar{ \delta \tilde{W}_{\mathrm{u}}}$.  Figure \ref{fig:workandunitary} displays $\delta \tilde{W}$ for a single quantum trajectory in the presence of feedback (blue) and for a different trajectory in the absence of feedback (red) compared to the expected unitary work $\ar{ \delta \tilde{W}_{\mathrm{u}}}$ (green).  Figure  \ref{fig:workandunitary}b,c show that in the presence of feedback the work is more closely correlated with the unitary work, with the correlation only limited by the efficiency of the feedback loop \kwm{\cite{sm}}. \el{In the limit of unit quantum efficiency and null loop delay, a feedback loop could exactly compensate for the exchanged heat \cite{sm}.}

\textit{Conclusion.} We have introduced experimental protocols for a continuously monitored driven  qubit that allow  to operationally define and individually measure quantum heat and work along single trajectories,  \kwm{accounting for the presence of coherent  superpositions of energy eigenstates. We have verified the first law of thermodynamics at the level of energy exchanges and of transition probabilities. Moreover, we have demonstrated the consistency of these results with the master equation approach as well as with the TPM scheme, both for open and closed evolutions, with the help of feedback control.}
Our findings pave the way for  future experimental and theoretical studies  in quantum  thermodynamics  \cite{gem04} at the single trajectory level.

\begin{acknowledgements}
 \textit{Acknowledgements:} We acknowledge research support from the NSF (Grants No. PHY-1607156 and No. PHY-1752844 (CAREER)), the ONR  (\el{G}rant No. 12114811), the John Templeton Foundation,  and the EPSRC (Grant No. EP/P030815/1). This research used facilities at the Institute of Materials Science and Engineering at Washington University. K.~W.~M. acknowledges support from the Sloan Foundation. \el{E.~L. acknowledges support from the German Science Foundation (DFG) (Grant No.~FOR 2724).}

\end{acknowledgements}

%\bibliographystyle{naturemag}
%\bibliographystyle{unsrt_withcaps}
%\bibliographystyle{apsrev}

%\bibliography{ppp_references-1-1}


\widetext

\section{Supplemental Material}
 
\subsection{Heat and work definitions and contributions to transition probabilities}

{For a continuously monitored driven quantum system, one can consistently identify unitary and nonunitary contributions to (i) the evolution of the conditional density operator $\tilde{\rho}_t$ along a single  trajectory,  (ii) to the energy of the system $\tilde U_t$ (averaged over quantum fluctuations), and (iii) to changes $\delta \tilde{P}_{m,n}$ of the transition probabilities \cite{alon16}}.
The definition of heat and work in the manuscript stems from the association of work with the deterministic driving of the quantum system, and of heat with the stochastic evolution due to the detection process, $d \tilde{\rho}_t= \delta \mathbb{W}[\tilde \rho_t] dt+ \delta \mathbb{Q}[\tilde \rho_t]dt$, {where  $\delta \mathbb{W}[\tilde \rho_t]$ and $\delta \mathbb{Q}[\tilde \rho_t] $ are  superoperators associated with the respective unitary and nonunitary dynamics.}
In fact, the change in the internal energy between times $t$ and $t+dt$ along a single quantum trajectory may be    expressed as,
\kwm{\begin{align}
\label{intener}
d \tilde U_t=& \tr[H_t(\tilde\rho_{t-dt}+d\tilde\rho_t)]-\tr[H_{t-dt}\tilde\rho_{t-dt}]\nonumber\\
=& \tr[ \tilde\rho_{t-dt}dH_t]+\tr[H_{t}\delta\mathbb{W}[\tilde \rho_t]dt ] + \tr[H_{t}\delta\mathbb{Q}[\tilde \rho_t]dt ]\nonumber\\
=&\delta \tilde W_t + \delta \tilde Q_t,
\end{align} }
where in the second line $dH_t=H_t-H_{t-dt}$ and  \kwm{$\tr[H_t\delta\mathbb{W}[\tilde \rho_t]dt]= -({i}/{\hbar}) \text{tr}[H_t [H_t,\rho_t] dt]=0$.} In the last line, $\delta\tilde W_t=\tr[\tilde\rho_{t-dt}dH_t]$ and $\delta\tilde Q=\tr[H_{t}d\tilde{\rho}_t]$, indicating that work is related to a change of the Hamiltonian and heat to the nonunitary changes in the state. {Equation \eqref{intener} shows that there are actually two contributions to the averaged work $\delta \tilde W_t$: one coming from the variation $dH_t$ of the Hamiltonian and one coming from the superoperator  $\delta \mathbb{W}[\tilde \rho_t]$. However, the average of the latter vanishes due to the properties of the trace. The superoperator  $\delta \mathbb{W}[\tilde \rho_t] $, by contrast, directly contributes to the (unaveraged) density operator $\tilde{\rho}_t$ and transition probabilities $\delta \tilde{P}_{m,n}$. We note that the association of $\delta \mathbb{W}[\tilde \rho_t] $ with work at this level is limited to the case of driven unitary dynamics.}

The different contributions of heat and work to the quantum evolution are reflected in different contributions to the transition probabilities.
The changes to the transition probabilities due to heat and work are defined as $\delta \tilde{P}_{m,n}^Q = \tr\left[\Pi_m\,\dbar\mathbb{Q}[\tilde{\rho}_t]\right]$, and $\delta \tilde{P}_{m,n}^W =  \tr\left[\Pi_m\,\dbar\mathbb{W}[\rho_t]\right]$,  where $\Pi_m$ is the projective measurement operator of the eigenstate $m$ at time $t$  and the trajectory $\tilde{\rho}(t)$ originates in eigenstate $n$. 
Correspondingly, we define the path-dependent total transition probabilities, 
\begin{align}\label{eq:probb}
\tilde{P}_{m,n}^Q = \int_0^\tau dt\, \delta \tilde{P}_{m,n}^Q,\quad
\tilde{P}_{m,n}^W = \int_0^\tau dt\, \delta \tilde{P}_{m,n}^W .
\end{align}
The definition of heat and work contributions to the evolution of the density matrix, $d \tilde{\rho}_t$, imply that, starting from a density matrix $\rho_0$ corresponding to an eigenstate $n$, the total transition probabilities $\tilde{P}_{m,n}^\tau$ are given by \addkm{
\begin{equation}
\begin{split}
\tilde{P}_{m,n}^\tau  = \tr[\Pi_m \tilde{\rho}_t] = P^0_{m,n}+ \int_0^\tau dt\, \tr[\Pi_m \delta \mathbb{Q}[\tilde \rho_t]]dt 
 + \int_0^\tau dt\, \tr[\Pi_m \delta \mathbb{Q}[\tilde \rho_t]]dt = \tilde{P}_{m,n}^Q+ \tilde{P}_{m,n}^W,
\end{split} \label{eq:traiettoria}
\end{equation}}
\addkm{The instantaneous heat, work, and feedback are also expressed in terms of energy by $\dbar \tilde Q =\hbar \omega_\mathrm{q}\mathrm{tr}\left[\Pi_{m=1} \dbar\mathbb{Q}[\tilde \rho_t]\right]$, $\dbar \tilde W =\hbar \omega_\mathrm{q}\mathrm{tr}\left[\Pi_{m=1} \dbar\mathbb{W}[\tilde \rho_t]\right]$, and $\dbar \tilde W_\mathrm{F} =\hbar \omega_\mathrm{q}\mathrm{tr}\left[\Pi_{m=1} \dbar\mathbb{W}_\mathrm{F}[\tilde \rho_t]\right]$, respectively. }These changes reflect only the instantaneous changes in energy and do not depend on the initial state of the system.

Transition probabilities can be experimentally obtained by preparing the qubit in a specific initial state $n$ and terminating the experiment at time $\tau$ with a projective measurement  \cite{reed10}
 in the energy basis. These projective measurements allow us to determine the total \kwm{energy change} the qubit and compare to the total energy calculated from the heat and work along individual quantum trajectories, in a manner similar to the tomographic validation for trajectories in previous work  \cite{murc13traj,webe14}.  
  In Figure 1d we show the total energy changes of the qubit $\Delta U$ obtained from the transition probability $P_{1,0}$ which was determined from an ensemble of $10^5$ experiments of variable duration $\tau$ where the path-dependent energy $\tilde{U}$ was within $\pm 0.05\ \hbar \omega_\mathrm{q}$ of the black curve shown in Figure 1d at time $\tau$.  The transition probability is corrected for the finite readout fidelity of 65\%.  In Figure 1e, we compare the measured, path-independent, transition probability $P_{0,0}$, to the path-dependent transition probability $\tilde{P}^W_{0,0} + \tilde{P}^Q_{0,0}$, by binning experiments of variable duration according to the final path-dependent transition probability and determining the transition probability $P_{0,0}$ at each point from the outcomes of the projective readout. The close agreement between  $P_{0,0}$ and $\tilde{P}^W_{0,0} + \tilde{P}^Q_{0,0}$ indicates that the two independent measures of the qubit energy are in agreement, thereby confirming the first law of thermodynamics. 

\subsection{The unitary limit}

\kwm{In the absence of a detector, the system under consideration reduces to an isolated system and our definition \kwm{of} work agrees with the standard definition in Eq. (1) of the manuscript. In order to see this for a two energy measurement protocol,  we note that our distinction between heat and work amounts to the separation of work-like and heat-like components of the transition probabilities, $\tilde{P}^W_{m,n}$ and $\tilde{P}^Q_{m,n}$. When the system detector coupling is vanishing,  one has $\tilde{P}^W_{m,n} = P_{m,n}^{\tau}$. In fact, for an isolated system initially prepared in the $E_n^0$ energy eigenstate, the trajectory is only dictated by the unitary evolution, with the nonunitary part being zero along the entire trajectory, $ \delta \mathbb{Q}[\tilde \rho_t]/dt =0$. Hence $\tilde{P}^Q_{m,n}=0$ and $d \tilde{\rho}_t= \delta \mathbb{W}[\tilde \rho_t] dt$. Together with Eq. \eqref{eq:traiettoria}, this implies the work-like component of the transition probability is reduced to that of an isolated system, 
\begin{equation}
\tilde{P}^W_{m,n}=\tilde{P}^{\tau}_{m,n}={\rm tr} [ \Pi_m \tilde{\rho}_{\tau} ]=P^{\tau}_{m,n}.
\end{equation} }

%Our formalism reproduces the physics of the isolated system also for the average quantities computed along single trajectories, $\delta \tilde{W}$ and $\delta \tilde{Q}$.  In fact, for an isolated system, $ \delta \mathbb{Q}[\tilde \rho_t]=0$, hence $\delta \tilde{Q}/dt=0$. We can therefore write Eq. (2) in the manuscript as 
%%\begin{equation}
%%\Delta U= \int_0^{\tau}\frac{\delta W}{dt} dt =\sum_{j=1}^N {\rm tr} [\tilde{\rho_{t_{j-1}} (H_{t_j}-H_{t_{j-1}})]={\rm tr} [\rho_{t_j}H_{t_j} - \rho_{t_0}H_{t_0}] - \sum_{j=1}^N {\rm tr} [H_{t_{j}}(\rho_{t_j}-\rho_{t_{j-1}})] = {\rm tr} [\rho_{t_j}H_{t_j} - \rho_{t_0}H_{t_0}] -
%%\end{equation}
%\begin{equation}
%\begin{split}
%\Delta \tilde{U}= \int_0^{\tau}\frac{\delta W}{dt} dt = \int_0 ^{\tau} {\rm tr} [ \rho_t \frac{dH}{dt}]dt={\rm tr} [\rho_{\tau} H_{\tau} - \rho_0 H_0] -{\rm tr}\int_0^{\tau} [\frac{d \rho_t}{dt} H] dt \\ = {\rm tr} [\rho_{\tau} H_{\tau} - \rho_0 H_0] + i \int_0^{\tau} {\rm tr}[[H,\rho_t] H] dt =  {\rm tr} [\rho_{\tau} H_{\tau} - \rho_0 H_0] = {\color{blue} \langle  \Delta {U} \rangle},
%\end{split}
%\end{equation}
%where the average in the last equality is that obtained from the distribution in Eq.~(1) of the manuscript. {\color{blue} We have here used that $\tilde \rho_{t-dt} \rightarrow \tilde\rho_{t}$ in the limit $dt \rightarrow 0$.}

In addition, our formalism also reproduces the physics of an isolated system at the level of a single two-energy-measurement realization. \kwm{In that case, one identifies four possible trajectories corresponding to the transitions from the initial states of energy $E_n^0$ to the final states with energy $E_m^{\tau}$. The probability of such a trajectory is $P_n^0 P_{m,n}^{\tau}$. Each of these trajectories consists of a unitary evolution from $0$ to $t_f$ ending with a density matrix $\rho_f$ and a final extra  nonunitary step from $t_f$ to $t_f+\Delta t_M=\tau$ }determined by the measurement, \aleadd{during which the Hamiltonian is unchanged} and $\rho_f \to \rho_m=\vert m \rangle \langle m \vert$.  
%Since the extra measurement step is a purely nonunitary contribution, we can compute 
 The work contribution from the unitary evolution is,
\begin{eqnarray}
\int_{0}^{\tau} \frac{\delta  \tilde W}{dt} \, dt= \int_{0}^{t_f} \frac{\delta \tilde  W}{dt} \, dt &=& {\rm tr} [\tilde  \rho_f H_{\tau} - \tilde \rho_0 H_0] \nonumber \\&=& {\rm tr} [\tilde \rho_f   H_{\tau} ] - E_n^0 .
\end{eqnarray}
\kwm{The contribution from the last step does not involve exchange of energy with the detector and is therefore regarded as work, although it arises from nonunitary evolution~\cite{horo12}. This yields}
\begin{align}
\int_{0}^{\tau} \frac{\delta \tilde  W}{dt} \, dt&= \int_{t_f}^{t_f+\Delta t_M} \frac{\delta \tilde  W}{dt} \, dt & \nonumber\\   &= {\rm tr} [\tilde  \rho_m H_{\tau} - \tilde  \rho_f H_{\tau}] = E_m^{\tau}- {\rm tr} [\tilde  \rho_f   H_{\tau} ] .
\end{align}
This correctly reproduces the change of internal energy, $\Delta \tilde U = \Delta U = E_m^\tau -E_n^0= W$ associated with the trajectory from energies $E_n^0$ to $E_m^\tau$.  We therefore recover the full probability distribution in Eq. (1) of the manuscript.

%\textbf{Homodyne detection}---The qubit is realized by the near-resonant interaction of a transmon circuit   \cite{koch07}
% and a three-dimensional aluminum cavity  \cite{paik113D}
%  which is coupled to a $50\  \Omega$ transmission line. The radiative interaction between the qubit and transmission line is given by the interaction Hamiltonian, $H_\mathrm{int.}=\hbar\gamma(a \sigma_+ + a^{\dagger} \sigma_ -)$, where $\gamma$ is the coupling rate between the electromagnetic field mode corresponding to $a$ ($a^\dagger$), the annihilation (creation) operator, and qubit state transitions by $\sigma_+$ ($\sigma_-$), the raising (lowering) ladder operator for the qubit. By virtue of this interaction Hamiltonian, a homodyne measurement along an arbitrary quadrature of the quantized electromagnetic field of the transmission line, $a e^{-\phi} + a^{\dagger} e^{+\phi}$, results in weak measurement along the corresponding dipole of the qubit, $\sigma_+ e^{-\phi} + \sigma_- e^{+\phi}$   \cite{nagh16}.
%   Homodyne monitoring is performed with a  Josephson parametric amplifier \cite{cast08,hatr11para}   operated in phase-sensitive mode. We adjust the homodyne detection quadrature such that the homodyne signal $dV_t$ obtained over the time interval $(t, t+dt)$ provides an indirect signature  \cite{jord15}     of the real part of $\sigma_- = (\sigma_x+i\sigma_y)/2$. The detector signal  is given by $d V_t = \sqrt{\eta}\gamma \langle \sigma_x \rangle dt + \sqrt{\gamma} dX_t$, where $dX_t$ is a zero-mean Gaussian random variable that arises due to the quantum fluctuations of the field mode.

\subsection{Heat and work tracking}

The stochastic master equation (Eq. 3) is used to update state of the qubit conditioned on the collected homodyne signal which is digitized in $20$ ns steps, and scaled such that its variance is $\gamma dt$ \cite{nagh16}.  Our identification of work and heat as the respective unitary and nonunitary changes of the state applies in the laboratory frame. However, it is convenient to calculate the state trajectories in the frame rotating with the qubit drive and identify energy changes in the rotating frame. \kwm{We break the evolution into discrete timesteps. Each timestep $i$ is divided into two substeps. The first substep updates the $\tilde{\rho}[i]$ by the unitary terms.  The second substep updates $\tilde{\rho}[i]$ with the non-unitary terms given by the discretized stochastic master equation (in It$\hat{\mathrm{o}}$ form) \cite{alon16,nagh16caustic}. }
\begin{align}
d\tilde{\rho}_{00}[i] =      \gamma(1-\tilde{\rho}_{00}[i]) dt + \sqrt{\eta}(dV[i]-\sqrt{\eta} \gamma 2 \tilde{\rho}_{01}[i] dt)  2\tilde{\rho}_{01}[i] (1-\tilde{\rho}_{00}[i]),\\
d \tilde{\rho}_{01}[i] = \gamma \tilde{\rho}_{01}[i]/2 dt + \sqrt{\eta}(dV[i]-\sqrt{\eta}\gamma 2 \tilde{\rho}_{01}[i]dt) (1-\tilde{\rho}_{00}[i]-2\tilde{\rho}^2_{01}[i]).
\end{align}  \label{sme:ito}
%\begin{align}
%d\tilde{\rho}_{00}[i] =    \Omega_\mathrm{R} \tilde{\rho}_{01}[i] dt +  \gamma(1-\tilde{\rho}_{00}[i]) dt + \sqrt{\eta}(dV[i]-\sqrt{\eta} \gamma 2 \tilde{\rho}_{01}[i] dt)  2\tilde{\rho}_{01}[i] (1-\tilde{\rho}_{00}[i]),\\
%d \tilde{\rho}_{01}[i] = \Omega_\mathrm{R} (1/2-\tilde{\rho}_{00}[i]) dt  -\gamma \tilde{\rho}_{01}[i]/2 dt + \sqrt{\eta}(dV[i]-\sqrt{\eta}\gamma 2 \tilde{\rho}_{01}[i]dt) (1-\tilde{\rho}_{00}[i]-2\tilde{\rho}^2_{01}[i]).
%\end{align}  \label{sme:ito}
 Therefore, in each timestep, we accordingly distinguish between instantaneous work and instantaneous heat in the rotating frame. \kwm{Since the transformation to the lab frame is set by the deterministic driving, it is  straightforward to determine these energy changes in the lab frame.}

\subsection{Feedback}

In this section, we analyze in detail the feedback protocol used in the experiment and compare it with optimal feedback protocols in the presence of finite efficiency and feedback loop delay.

\kwm{If $\eta=1$ and given a pure initial state, the state of the system is pure at  all times, and is described by a vector on the surface of the Bloch sphere. An ideal feedback loop would then exactly and immediately compensate for the heat exchanged with the environment resulting in completely unitary evolution of the state. 
This is not possible at finite inefficiency, where the evolution of the system is no longer constrained to the surface of the Bloch sphere. Since the evolution due to the feedback protocol is unitary, it preserves the length of the Bloch vector, and cannot maintain pure} evolution once purity has been lost. Therefore, it is impossible to exactly compensate for the exchanged heat with a unitary operation. %This means even if we have immediate access to $x$ and $z$ and able to respond immediately we would not perfectly close the system and recover full contrast unitary evolution. In reality, Feedback loop delay also reduces the efficiency of the feedback.\\

Given access to the trajectory in real time, the best possible feedback is to maintain the phase of the oscillation as if \kwm{the qubit state} were to undergo closed unitary evolution (Supplemental Fig.\ \ref{sub1}a). In this case, the feedback output would be 
$\Omega_\mathrm{F} dt=- \theta_Q $ where $\theta_Q$ can be calculated by the state of the qubit at $t$ and $t+dt$. In this case we would not have control over the purity of the state and it will change by measurement backaction from point to point. Supplemental Figure~\ref{sub1} shows simulation results for this type of feedback for $35\%$ quantum efficiency. As depicted in Supplemental Figure~\ref{sub1}b, the exchanged heat contribution is compensated by the feedback contribution. The scatter plot in Supplemental Figure \ref{sub1}(c) shows the anti-correlation of these contributions. Therefore, as we expect, the system will behave more like a closed system and we observe persistent Rabi oscillations with $70 \%$ of full contrast (Supplemental Fig.\ \ref{sub1}d). However, a realistic feedback loop would also have a finite loop delay, given by the time it takes for the measurement signal to travel to a detector and for the feedback output to be calculated. Considering a feedback loop delay of $\sim 500$ ns  \cite{camp13strob,Rist12},
 Supplemental Figure \ref{sub1}(d) (blue curve) shows that the feedback performance would be reduced.

\setcounter{figure}{0}

\begin{figure}
\renewcommand\figurename{Supplemental Figure}
  \begin{center}
    \includegraphics[width=0.9\textwidth]{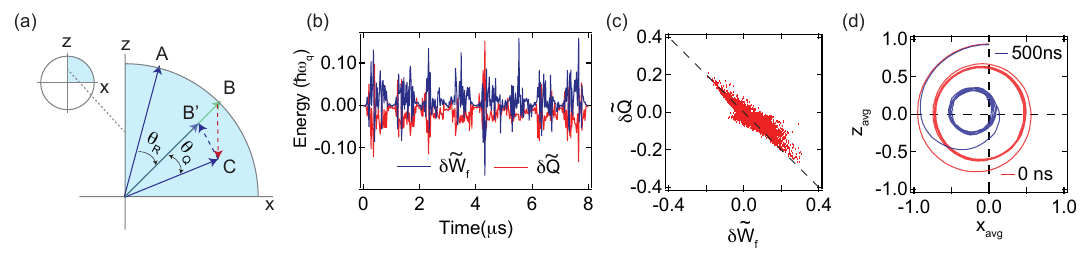}
  \end{center}
  \vspace{-.2in}
  \caption{\kwm{ \textbf{Simulation of optimal unitary feedback.} (a), Schematic of the feedback operation, the qubit evolution from $A$ at time $t$ to $C$ at time $t+dt$ comprises of two different types of evolution; unitary evolution due to the Rabi drive ($A \rightarrow B$) and stochastic evolution due to coupling to the environment. Because the environment is monitored with nonideal quantum efficiency we effectively average over some of the stochastic evolution reducing the state purity  ($B \rightarrow C$). The ideal unitary feedback maintains the phase relation with unitary evolution by application of a rotation $-\theta_Q$ to state $B'$. (b), Instantaneous contributions of heat and feedback to the transition probability $P_{0,0}$ for a single run of experiment. (c), Scatter plot of the instantaneous heat versus feedback for 100 runs of experiment which shows an anti-correlation ($r=-0.9$) between feedback and heat contributions to the transition probability $P_{00}$. (d), Ensemble behavior of the trajectories in presence of feedback with no delay (red) and a $500$ ns loop delay (blue) showing a persistent Rabi oscillations. }}\label{sub1}
\end{figure}

The feedback loop implemented in our experiment differs from the optimal feedback in that it does not require real-time state tracking and error processing.  This feedback takes a copy of homodyne signal,  $dV_t$, and multiplies it by a sinusoidal reference signal, $A\left[\sin(\Omega_R t + \phi)+ B\right]$ resulting in a feedback signal $\Omega_\mathrm{F}$, which is used to modulate the drive amplitude. %to compensate for heat in each timestep (In our experiment $A=1$v for $\Omega_R=1$MHz and $B=-1$v) The output of feedback loop then mixed with qubit frequency and added to the drive.

\begin{figure}
\renewcommand\figurename{Supplemental Figure}
  \begin{center}
    \includegraphics[width=0.9\textwidth]{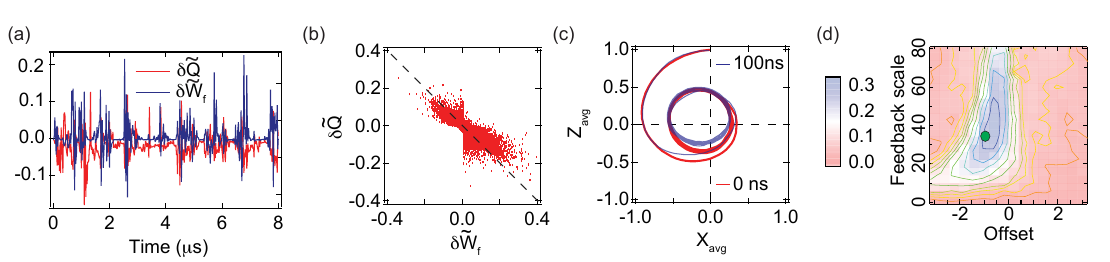}
  \end{center}
  \vspace{-.2in}
  \caption{ \textbf{Simulation of phase-lock feedback.} \small   (a), Instantaneous contribution of heat and feedback to the transition probability $P_{00}$ for a single run of experiment. (b), Scatter plot of instantaneous heat versus feedback for 100 runs of the experiment which shows anti-correlation  ($r=-0.81$) between the feedback and heat contributions to transition probability $P_{0,0}$. (c), Ensemble behavior of the trajectories in presence of feedback with no delay (red) and $100$ ns loop delay (blue) showing a persistent Rabi oscillations. (d), Simulated feedback efficiency versus feedback scale ($A$) and offset $(B)$.  The green dot indicates the expected parameters for optimal feedback.}\label{sub2}
\end{figure}

This feedback loop essentially implements a phase-locked loop, and in order to clarify how the loop works we may cast the stochastic master equation (3) in terms of the Bloch components $x$ and $z$,
\begin{align}\label{eq:smezz}
dz = + \Omega x dt+ \gamma (1-z) dt
+\sqrt{\eta} x (1-z) ( dV_t - \gamma \sqrt{\eta} x dt)\\
dx = -\Omega z dt -\frac{\gamma}{2} x dt 
+\sqrt{\eta} (1-z - x^2)(dV_t -\gamma \sqrt{\eta} x dt ) .
\end{align}
\kwm{It is apparent that by canceling the last two terms, the evolution would be  unitary as we expect for} a closed system. However, with only unitary rotations we can change $dz$ and $dx$ in the following way,
\begin{eqnarray}
dz = \Omega_\mathrm{F} x dt ,\quad 
 dx = -\Omega_\mathrm{F} z dt, \label{xf}
\end{eqnarray}
where, we wish to cancel all the stochastic terms in (\ref{eq:smezz}) with the unitary terms (\ref{xf}).
%\begin{subequations}
%\begin{eqnarray}
%\Omega_f x dt &\nleftrightarrow &  + \gamma (1-z) dt
%+\sqrt{\eta} x (1-z) ( dI_t - \gamma \sqrt{\eta} x dt), \label{cancelz}\\
% -\Omega_f z dt & \nleftrightarrow &  -\frac{\gamma}{2} x dt 
%+\sqrt{\eta} (1-z - x^2)(dI_t -\gamma \sqrt{\eta} x dt ). \label{cancelx}
%\end{eqnarray}
%\end{subequations}
\kwm{Regardless how complicated $\Omega_\mathrm{F}$ is, with finite efficiency, it is impossible to compensate for all terms as mentioned earlier and the best choice recovers about $70 \%$ of purity for 35\% quantum efficiency. To understand how the phase-locked loop approximates the optimal feedback, we consider just the $z\equiv\langle\sigma_z\rangle$} component of the state. This is reasonable since all thermodynamics parameters e.g. work, heat and transition probabilities directly relate to the $z$ component. This requires $\Omega_\mathrm{F} x dt = -\sqrt{\eta} x (1-z) ( dV_t - \gamma \sqrt{\eta} x dt)$, where for weak measurement $\gamma \sqrt{\eta} x dt$ is negligible compared to $dV_t$.  Thus, we have  $\Omega_\mathrm{F}  = - \sqrt{\eta} (1-z) dV_t/dt $. The essence of the phase-locked loop is to replace $z$ with $\cos(\Omega t + \phi)$, which is the ``target" $z$ that would be obtained for closed evolution. Here $\phi=0\ (\pi)$ for an initial ground (excited) state. This choice for $z$ has a \kwm{two-fold effect: not only is this a reasonable approximation} for $z$ in presence of feedback but it also locks the  oscillation phase which addresses the damping term in (\ref{eq:smezz}). Note that the choice of phase $\phi$ only affects the transient behavior and appears as a overall phase shift in the persistent Rabi oscillations without affecting the contrast. Thus the feedback signal is,
\begin{subequations}
\begin{eqnarray}
\Omega_\mathrm{F}  =  \sqrt{\eta} (\cos(\Omega t + \phi)-1) dV_t/dt , \label{jcancelz}
\end{eqnarray}
\end{subequations}
This equation suggests that the scale for feedback should be around $A=\sqrt{\eta}/dt \sim 30$ which is in agreement with optimal value found empirically in the experiment of $A=34$. Experimentally, this factor may be set by pre-amplification of the homodyne signal. Note that this result also suggests the offset term of $B=-1$ as we use in our feedback setup. Supplemental Figure~\ref{sub2} shows the simulation result for the phase-locked feedback with $35 \%$ quantum efficiency.

As we see in Supplemental Figure~\ref{sub2}(a,b), the phase-locked feedback loop effectively compensates for the heat at each point in time. Supplemental Figure~\ref{sub2}(c) shows persistent Rabi oscillations for this case. In Supplemental Figure \ref{sub2}(d), we explore the contrast of persistent Rabi oscillations versus feedback parameters. The simulated result shows that maximum contrast occurs around $B \sim -1$ and $A \sim 35$ as we expect.

\subsection{Experimental setup and parameters} 

The transmon circuit was fabricated by double angle evaporation of aluminum on a high resistivity silicon substrate. The circuit was placed at the center of a 3D aluminum waveguide cavity machined from 6061 aluminum. The bare cavity frequency is $\omega_\mathrm{c}/2\pi=7.257$ GHz. The near-resonant interaction between the circuit and the cavity (characterized by coupling rate $g/2 \pi = 136$ MHz) results in hybrid states, as described by the Jaynes-Cummings Hamiltonian. The lowest energy transition of hybrid states ($\omega_\mathrm{q}/2\pi=6.541$ GHz) can therefore be considered a ``one-dimensional" artificial atom because the radiative decay of the system is dominated by the cavity's coupling to a  50 $\Omega$ transmission line.  This radiative decay was characterized by a decay of rate $\gamma=1.7$ $\mu$s$^{-1}$. Resonance fluorescence  from the artificial atom is amplified by a near-quantum-limited Josephson parametric amplifier, consisting of a $1.5$ pF capacitor, shunted by a Superconducting Quantum Interference Device (SQUID) composed of two $I_0 = 1\ \mu $A Josephson junctions.
The amplifier produces 20 dB of gain with an instantaneous 3-dB-bandwidth of 50 MHz. The quantum efficiency was measured to be 35\%. We drive the qubit by sending a resonant coherent signal via a weakly coupled transmission line, and the strength of the drive is characterized by a Rabi frequency of $\Omega/2\pi \simeq 1$ MHz. The total loop delay for the feedback is 100 ns. The initial state fidelity was limited by a 3\% thermal population of the excited state. 

\begin{figure}
\renewcommand\figurename{Supplemental Figure}
\begin{center}
\includegraphics[width =.5\textwidth]{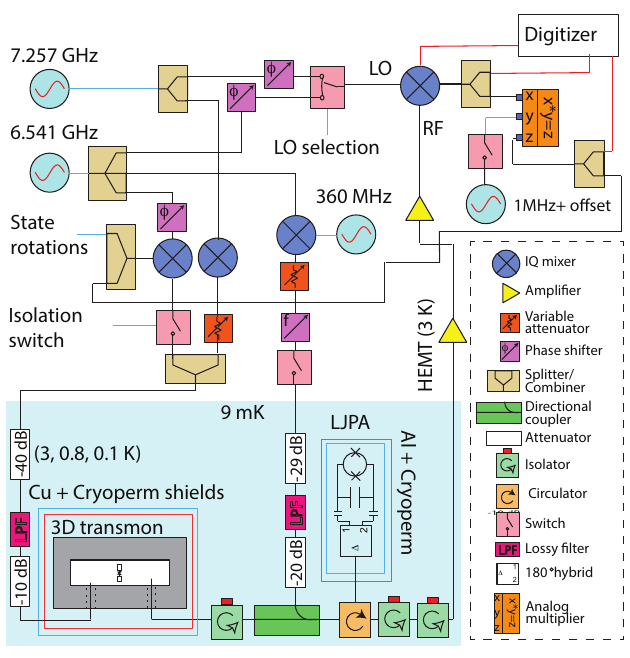}
\end{center}
\caption{\small {\bf Experimental setup} The qubit and Josephson parametric amplifier share a signal generator to maintain the phase relation defining the amplification quadrature. We use a double sideband technique to pump the parametric amplifier. The homodyne signal is split for the purpose of feedback and state tracking. Both the feedback signal and homodyne signal are digitized for state tracking in a post-processing step.}
\label{fig:schematic}
\end{figure}

\subsection{Statistical Analysis}

Error bars reported in Figure 1 indicate the standard error of the mean for binomial data, which are subsequently scaled by the readout fidelity.  On average 300, projective energy measurements were used for the determination of $\Delta U$ at each timestep in  Figure 1d. The number of data points for each value reported in Figure 1e is shown in the inset. In Figure 3, the shaded regions indicate the standard error of the mean based on binomial data used to determine the work distribution and the standard error of the mean based on the trajectories used to calculate the efficacy.

\end{document}